\documentclass[sigconf]{acmart}

\AtBeginDocument{%
  \providecommand\BibTeX{{%
    \normalfont B\kern-0.5em{\scshape i\kern-0.25em b}\kern-0.8em\TeX}}}

\copyrightyear{2020}
\acmYear{2020}
\setcopyright{acmcopyright}
\acmConference[SEAMS '20]{IEEE/ACM 15th International Symposium on Software Engineering for Adaptive and Self-Managing Systems}{October 7--8, 2020}{Seoul, Republic of Korea}
\acmBooktitle{IEEE/ACM 15th International Symposium on Software Engineering for Adaptive and Self-Managing Systems (SEAMS '20), October 7--8, 2020, Seoul, Republic of Korea}
\acmPrice{15.00}
\acmDOI{10.1145/3387939.3391604}
\acmISBN{978-1-4503-7962-5/20/05}

\usepackage[english]{babel}
\usepackage{graphicx}
\usepackage{amsmath,amssymb,amsfonts}
\usepackage{graphicx}
\usepackage{textcomp}
\usepackage{booktabs} 
\usepackage{pifont}
\usepackage{caption}
\usepackage{subcaption}
\usepackage[linesnumbered,ruled,vlined]{algorithm2e}
\usepackage{algpseudocode}
\usepackage{multirow}
\usepackage{textcomp}
\usepackage[listings,skins,breakable]{tcolorbox}
 
 \newcommand{\approach}{\texttt{DATESSO}}
 
\title{UploadingImages}

\graphicspath{{images/}}
\begin{document}

\title{DATESSO: Self-Adapting Service Composition with Debt-Aware Two Levels Constraint Reasoning}


\author{Satish Kumar}
\affiliation{%
  \institution{School of Computer Science}
   \institution{University of Birmingham, UK}}
\email{s.kumar.8@cs.bham.ac.uk}

\author{Tao Chen}
\affiliation{%
   \institution{Department of Computer Science}
   \institution{Loughborough University, UK}}
\email{t.t.chen@lboro.ac.uk}

\author{Rami Bahsoon}
\affiliation{%
  \institution{School of Computer Science}
   \institution{University of Birmingham, UK}}
\email{r.bahsoon@cs.bham.ac.uk}

\author{Rajkumar Buyya}
\affiliation{%
\institution{School of Computing and Information Systems}
  \institution{University of Melbourne, Australia}}
\email{rbuyya@unimelb.edu.au}


\begin{abstract}
The rapidly changing workload of service-based systems can easily cause under-/over-utilization on the component services, which can consequently affect the overall Quality of Service (QoS), such as latency. Self-adaptive services composition rectifies this problem, but poses several challenges: (i) the effectiveness of adaptation can deteriorate due to over-optimistic assumptions on the latency and utilization constraints, at both local and global levels; and (ii) the benefits brought by each composition plan is often short term and is not often designed for long-term benefits\textemdash a natural prerequisite for sustaining the system. To tackle these issues, we propose a two levels constraint reasoning framework for sustainable self-adaptive services composition, called \approach. In particular, \approach~consists of a refined formulation that differentiates the `strictness' for latency/utilization constraints in two levels. To strive for long-term benefits, \approach~leverages the concept of technical debt and time-series prediction to model the utility contribution of the component services in the composition. The approach embeds a debt-aware two level constraint reasoning algorithm in \approach~to improve the efficiency, effectiveness and sustainability of self-adaptive service composition. We evaluate \approach~on a service-based system with real-world WS-DREAM dataset and comparing it with other state-of-the-art approaches. The results demonstrate the superiority of \approach~over the others on the utilization, latency and running time whilst likely to be more sustainable.
\end{abstract}


\maketitle
\section{Introduction}

Service composition allows software to be built by seamlessly composing readily available service components, each of which offers different guarantee on Quality-of-Services (QoS), where latency can be of paramount importance~\cite{zeng2004qos}. Dynamically composing services is an enabling property for service-based systems supported by Cloud, Edge, Smart  and Internet-of-Things environments. However, a known difficulty in service-based systems is the presence of rapidly changing workload, leading to under-/over-utilization on the services components~\cite{kumar2019identifying}. On one hand, increasing workload can enhance the over-utilization of a services component within a composite service, which in turns, would negatively affect the latency and may violate the Service Level Agreement (SLA)~\cite{raimondi2008efficient}~\cite{kumar2019identifying}. On the other hand, decreasing workload may lead to under-utilization of the capacity of component services, reducing the revenue that should have been achieved as the infrastructural resources also impose monetary cost. To address those issues, self-adaptation on service composition is promising, but the adaptation needs to be effective while being efficient and render benefits over time (i.e., sustainable).


When reasoning about self-adaptation for service composition, there are often two levels of latency/utilization constraints: the local constraint that relates to the individual constituent services and the global one for the entire service composition. Both of them are critical, as they can affect what the alternative composition plans to be searched during the adaptation~\cite{rosenberg2009end}. However, existing work on self-adapting service composition often rely on over-optimistic assumptions, such that both local and global constraints are hard and can always be satisfied~\cite{wang2013constraint,ardagna2005global,laleh2017constraint,de2007selection,rosenberg2009end}. This can negatively influence the adaptation quality and efficiency, rendering lengthy reasoning process, especially when the given constraints are unrealistic/inappropriate. Further, the manifestation of strong assumptions may completely ignore the fact that certain composition plans may temporally violate the constraint, but are likely to create much larger benefits after a certain period of time.

Given the rapidly changing workload, it is important to ensure that each adaptation can be effective over a period of time and would avoid unnecessarily frequent adaptations. However, current work informing adaptation tends to render short-term benefits, i.e., the immediate improvement of a composition plan. These improvements, for example, can be in response to (predicted) latency constraint violation/undesired utilization~\cite{dai2009qos,yang2009performance,aschoff2011qos}. Additionally, immediate low utilization/high latency in the short term may not necessarily mean an undesired composition plan; in fact, it can be the source that stimulates largely increased benefit in the long term. For example, under-utilization could be desirable temporarily in order to prepared for a largely increased workload in the long term. Similarly, over-utilization may be acceptable in short time, as long as the workload is only a `spike' and the loss can be paid off by long-term benefits. As a result, despite that adapting with composition plan that has the best immediate improvement may lead to short-term advantages, it can easily create instability and hinder the possibility of achieving higher benefits in the long term.


To address the above challenges, we propose a framework that leverages \underline{\textbf{d}}ebt-\underline{\textbf{a}}ware \underline{\textbf{t}}wo l\underline{\textbf{e}}vels con\underline{\textbf{s}}traint reasoning for \underline{\textbf{s}}elf-adapt-ing service c\underline{\textbf{o}}mposition (hence called \approach). We show that \approach~can achieve better utilization/latency in the long term while being faster than state-of-the-art approaches, providing more sustainable self-adaptive service-based systems. In a nutshell, the major contributions of this paper are summarized as follows:

\begin{itemize}
    \item Instead of formalizing the constraints at both local and global levels as hard ones, we refine the global constraints as the soft ones. This has enabled us to tailor the reasoning process in self-adaptation and mitigate over-optimism.
    \item We propose temporal debt-aware utility, a new concept that extends from the technical debt metaphor, to model the long-term benefit contribution of possible component services that constitute to a composition plan. 
    \item Drawing on the above, we design an efficient two level constraint reasoning algorithm in \approach~that is debt-aware, and utilizes the different strictness of the two level constraints to reduce the search space.
    \item We evaluate \approach~on a commonly used service-based system~\cite{DBLP:journals/infsof/ChenLY19,DBLP:conf/gecco/0001LY18,DBLP:conf/icpads/KumarBCLB18} whose component services are derived from the real-world WS-DREAM dataset~\cite{zheng2012investigating} and under the FIFA98 workload trace~\cite{arlitt2000workload}. The results show that, in contrast to state-of-the-art approaches~\cite{ardagna2005global} \cite{dai2009qos,lin2010design}, \approach~achieves better utilization and latency while having smaller overhead, leading to more sustainable self-adaptation in service composition.
\end{itemize}


The remaining of the paper is organized as follows: Section~\ref{sec:pre} presents the background information of service composition, the constraints, technical debt and a running example of the issues. Section~\ref{sec:overview} shows an overview of \approach. Section~\ref{sec:twolevel} discusses our formalization of the two level constraints with different strictness. The temporal debt-aware utility model and the debt-aware two level reasoning algorithm are specified in Section~\ref{sec:debtmodel} and~\ref{sec:reasoning}, respectively. Then, we presents the experiment results in Section~\ref{sec:exp}, following by discussion of threats to validity in Section~\ref{sec:tov}. Section~\ref{sec:rw} compares \approach~with existing work and Section~\ref{sec:con} concludes the paper.

\section{Preliminaries}
\label{sec:pre}

\subsection{Self-Adaptation in Service Composition}

A service composition is a special software form that consists of a particular workflow of connected abstract services, denoted as $\{a_{1}, a_{2},..., a_{x}\}$. Each of these abstract services can be realized by using a readily available component service selected from the Internet. Typically, there could be multiple component services to be selected, and the $y$th component service for the $x$th abstract service is denoted as $c_{xy}$. Therefore the possible component services for the $x$th abstract service form a set, denoted as $\{c_{x1}, c_{x2}, ...\}$, each of which has different generic latency guarantee on its capacity. For example, $c_{xy}$ has a capacity to process 50 requests in 0.5 seconds.

In such a context, a SLA may be legally negotiated to ensure the performance of a service composition by contract. The most notable elements in the SLA are the constraints on the utilization of service capacity and the achieved latency level per request, which we will elaborate in the next section.

As the workload changes, at runtime, the goal of self-adaptation for service composition is to find the composition plan, $\{c_{11}, c_{23},...,$ $c_{xy}\}$, that improves utilization and latency so that they satisfy all the constraints for as long as possible.

\subsection{Constraints in Service Composition}

In service composition, constraints denote the stakeholders' expectation of the latency guarantee. Most commonly, a SLA can define these constraints by specifying the bound of the latency and utilization~\cite{de2007selection}. For example, a service's latency should not exceed 10s or the utilization is at least 0.7. Typically, there are two levels of constraints:


\begin{itemize}

\item \textbf{Global constraint:} The global constraint specifies the minimum expectation of latency/utilization for the entire service composition. It is often the most common requirement in a service-based systems~\cite{lin2010design}~\cite{ardagna2005global}.

\item \textbf{Local constraint:} The local constraints are specified for the latency/utilization on each abstract service\footnote{For latency, this constraint would be applied for each request.}. This is important, as each abstract services can be realized by the component service from different parties; any violation of the local constraint would in fact cause severe failure in the composition, leading to an outage ~\cite{de2007selection}~\cite{ardagna2005global}.

\end{itemize}

It is worth noting that, satisfying all local constraints does not necessarily mean that the global constraint can be satisfied, since each of the constraints is documented separately~\cite{wang2013constraint}

\begin{figure*}[t!]
    \centering
\includegraphics[width=0.8\textwidth]{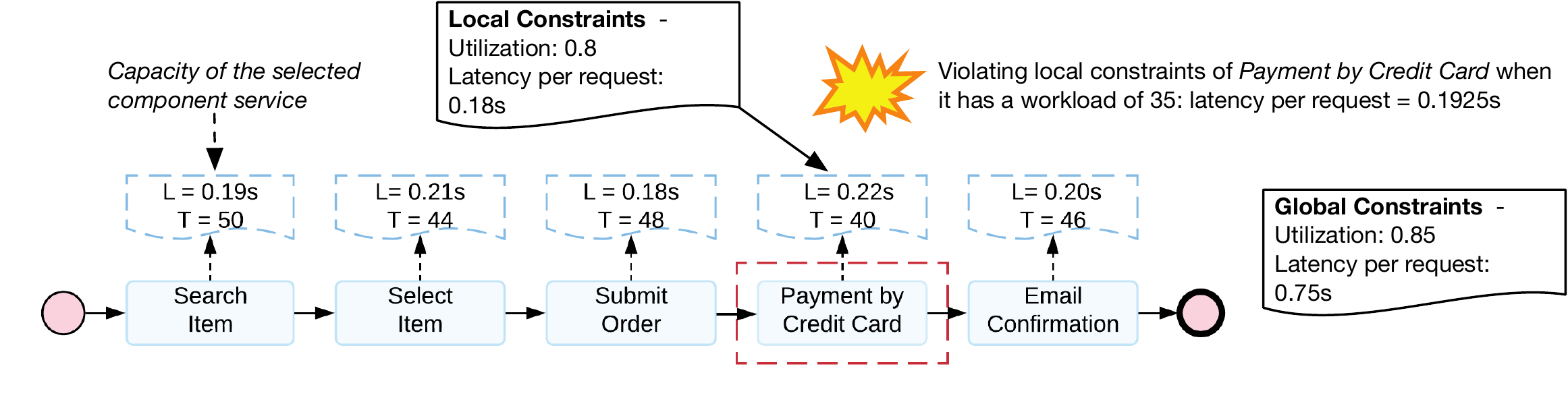}
    \caption{A running example of issues in service composition ($L$ and $T$ mean that the selected component service of an abstract service can process all $T$ requests in $L$ seconds)}
    \label{fig:MotExam}
\end{figure*}

\subsection{Technical Debt}
Technical debt is a widely recognized metaphor in software development~\cite{tom2013exploration,avgeriou2016managing,alves2016identification}. Its core idea is to describe the extra cost incurred by actions that compromise long-term benefits of the developed software, e.g., maintainability for short-term gains due to the need of timely software release.

The technical debt metaphor was initially introduced by Cunningham~\cite{cunningham1992wycash} in the context of agile software development, where the definition is described as:

\begin{tcolorbox}[breakable,title after break=,height fixed for=none,colback=blue!20!white,boxrule=0pt,sharpish corners,top=0pt,bottom=0pt,left=2pt,right=2pt]
\textit{``Shipping first-time code is like going into debt. A little debt speeds development so long as it is paid back promptly with a rewrite. The danger occurs when the debt is not repaid. Every minute spent on not quite right code counts as interest on that debt."} 
\end{tcolorbox}

In this regards, technical debt is often used in an economic-driven decision approach for communicating the technical trade-off between short-term advantages and long-term benefits in software projects ~\cite{tom2013exploration}. In the context of service composition, the notion of technical debt can be perfectly aligned with the requirement of long-term benefits: each possible component service may associate with a debt with respect to constraint violation. Such a debt, once selected, may or may not be repaid over a period of time, depending on the actual workload.

\subsection{Running Example}

In this section, we present a simple example of service composition to explain the problems. As shown in Figure~\ref{fig:MotExam}, there is a service composition in the form of sequentially connected abstract service, each of which has been realized by a particular component service. In this case, each selected component service has its own overall capacity, e.g., the selected component service for \textsc{Search Item} abstract service can process all 50 requests in 0.19 seconds.

As mentioned, each abstract service, along with the entire service composition, are legally documented with separated constraints on the utilization and latency per request, as specified in the SLA. Suppose that in this scenario, the local constraint of utilization and latency of each request for the abstract service \textsc{Payment by Credit Card} could be 0.8 and 0.18 seconds, respectively. Meanwhile, the global constraint of utilization and latency of each request for the service composition is 0.85 and 0.75 seconds, respectively. Given the changing workload, it is likely that either (or both) levels of constraint may be violated, which requires self-adaptation to replace the component services. However, there are two issues with this:

\begin{enumerate}

\item In this context, the different constraints are negotiated independently to each others. While it is relatively easy to find the alternative component service that satisfy the local constraints, searching for the composition plan that satisfies the global constraints is difficult, or we may not know whether one exists. As a result, existing approaches that treats both levels of constraints as hard constraints suffers the issue of being over-optimistic: they may struggle to find a satisfactory composition plan, especially under a scenario where such a plan barely exists. Further, this would completely eliminate the composition plan that may cause temporary violation of the global constraint(s), but can create much larger long-term benefits.

\item When self-adaptation is required, a possible component service and the entire composition plan may provide short-term immediate benefit in relieving constraint violation, but it is difficult to know whether such a benefit can be sustainable. In contrast, it is possible to temporally accept a composition plan that may still violate the global constraint(s), but will generate larger benefit in the long term. Therefore, self-adapting service composition without having any guarantee on the long term can lead to frequent adaptations with merely short-term benefits, which generate unnecessary overhead.
\end{enumerate}
The \approach~proposed in this work was designed to explicitly address these two issues in self-adapting service composition.


\section{\approach~Overview}
\label{sec:overview}

\begin{figure}[t!]
    \centering
\includegraphics[width=\columnwidth]{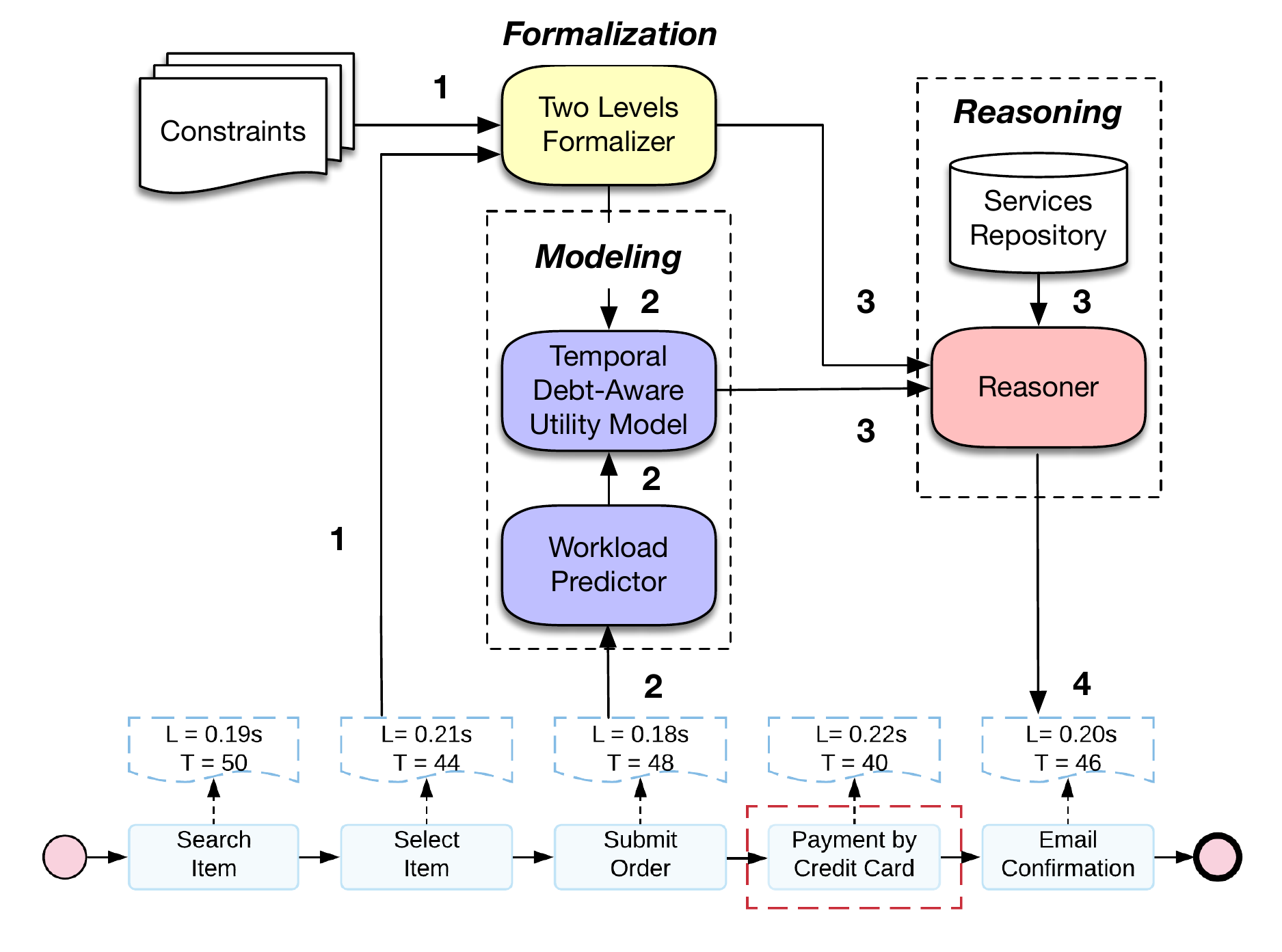}
    \caption{The general processes in \approach}
    \label{fig:arch}
\end{figure}

Figure~\ref{fig:arch} illustrates the overview of \approach. As can bee seen, there are three key stages, namely \emph{Formalization}, \emph{Modeling} and \emph{Reasoning}, each of which is specified as follows:

\begin{enumerate}
    \item \emph{\underline{Formalization:}} This is the very first stage in \approach~and it relies on the \emph{Two Levels Formalizer} component. Generally, it has two tasks at step 1: (i) formulating and recording the global/local level constraints as documented in the SLA; (ii) monitoring the service composition and informing the \emph{Modeling} stage, along with any information of the constraints, when any violations are detected. More details are discussed in Section~\ref{sec:twolevel}. Note that here, we trigger adaptation only based on local constraint violations, as we formalize the global ones as soft constraints. However, the global constraint is implicitly considered in the \emph{Reasoning} stage.
    \item \emph{\underline{Modeling:}} Once the local constraint violation has been detected, at step 2, the \emph{Workload Predictor} keeps track of the historical workload on each abstract service, and provides a time-series model to be embedded with the constraint information, which together form the temporal debt-aware utility model. A detailed discussion will be presented in Section~\ref{sec:debtmodel}
    \item \emph{\underline{Reasoning:}} At the final stage, the utility model that is debt-aware, the two level constraints and the \emph{Service Repository} with all possible component services would be exploited by the \emph{Reasoner} at step 3. Specifically, we design a debt-aware two levels constraint reasoning algorithm that (i) enables more efficient processing by reducing the original search space based on the constraint information, and (ii) produces a composition plan that is likely to have the highest long-term benefit, without explicitly using global constraints as caps or thresholds. Such a composition plan would then be sent for execution (step 4). The algorithm will be illustrated in greater details at Section~\ref{sec:reasoning}.
\end{enumerate}

Indeed, the components in \approach~can be formulated with a MAPE loop of self-adaptation~\cite{computing2006architectural}, but we did not explicitly perform such in this work for the purpose of better generality. In fact, \approach~is agnostic to the concrete architectural pattern, providing that the patterns meet with the needs of the components.

\section{Two Levels Constraints with different strictness}
\label{sec:twolevel}
As mentioned, we consider both local and global constraints for latency/utilization in the \emph{Formalization} stage of \approach. Instead of assuming hard constraint for both of them, we treat the global constraint as a soft one, which helps to mitigate the problem of being over-optimistic. The formal model and strictness of the two level constraints are discussed in the following subsections.

For each level, constraint can be related to both utilization and latency values. The utilization is a direct measurement of under-utilized situation, whilst the latency value reflects the problem of over-utilization, as a too high utilization usually means the component service is over-stressed, which results in latency degradation. 

\subsection{Hard Local Constraints}

As discussed in Section~\ref{sec:pre}, the local constraint is usually hard~\cite{ardagna2005global,alrifai2009combining}, which should not be violated. This is because at the service level, any violation of the constraint would in fact cause severe failure in the workflow execution. For example, a violation of latency/utilization caused by a workload that exceeds the capacity would simply bring the individual service down, which cause outage of the entire service composition.

Locally, for each component service $c_{xy}$ that has a capacity to process $T_{c_{xy}}$ requests in $L_{c_{xy}}$ seconds, we model the normalized constraint ($\varmathbb{CL}_{c_{xy}}$) on the normalized actual latency of each request ($\varmathbb{L}_{c_{xy}}$) to be satisfied as below, both of which are within $[0,1]$\footnote{Normalization can be achieved by using the lower and upper bounds of possible latency values.}:

\begin{gather}
\varmathbb{L}_{c_{xy}}={{L_{c_{xy}} \times W_{t,c_{xy}}} \over {T_{c_{xy}}}} \leq \varmathbb{CL}_{c_{xy}}
\end{gather}

\noindent where $W_{t,c_{xy}}$ is the workload for the corresponding abstract service (hence for $c_{xy}$ too) at timestep $t$. Likewise, the local constraint ($\varmathbb{CU}_{c_{xy}}$) on utilization ($\varmathbb{U}_{c_{xy}}$) to be satisfied can be formulated as\footnote{Utilization naturally sits within $[0,1]$, as any requests go beyond the capacity would be discarded.}:
\begin{gather}
\varmathbb{U}_{c_{xy}}={{L_{c_{xy}} \times W_{t,c_{xy}}} \over {\varmathbb{CL}_{c_{xy}} \times T_{c_{xy}}}} \geq \varmathbb{CU}_{c_{xy}}
\end{gather}

\noindent Since the local constraints are hard, we say a component service as \emph{feasible} if, and only if, both utilization and latency constraints are satisfied. Otherwise it is termed \emph{infeasible}.

\subsection{Soft Global Constraints}

Unlike existing work that model global constraint as hard threshold, we model its soft version that can tolerate certain violation, with an aim to mitigate the issue of over-optimism.
Indeed, the way of aggregating the local latency toward the global value for the entire service composition depends on the connectors, which may be sequential, parallel or recursive etc. However, as shown in~\cite{DBLP:journals/tweb/YuZL07,DBLP:journals/tweb/AlrifaiRN12}, sequential connector is the most fundamental type and all other connectors can be converted into a sequential one. Therefore in this work, we focus on sequential connector in our models.

Similar to its local counterpart, for all selected component services in the entire service composition, the satisfaction on normalized global constraint ($\varmathbb{CL}_{global} \in [0,1]$) and the normalized actual latency of each request ($\varmathbb{L}_{global} \in [0,1]$) can be calculated by aggregating the locally achieved latency. Specifically, when all the connectors are sequential or they have been converted into sequential ones, the satisfaction of global latency can be formulated as\footnote{We use $\preceq$ to reflect the `soft' nature of global constraints.}:

\begin{gather}
\label{eq:globallatency}
\varmathbb{L}_{global}=\sum_{x}\sum_{y} \varmathbb{L}_{c_{xy}} \preceq \varmathbb{CL}_{global}
\end{gather}


\noindent Likewise, the global constraint ($\varmathbb{CU}_{global}$) on utilization ($\varmathbb{U}_{global}$) to be satisfied can be formulated as:
\begin{gather}
\label{eq:globalutil}
\varmathbb{U}_{global}={1 \over N} \times \sum_{x}\sum_{y} \varmathbb{U}_{c_{xy}} \succeq \varmathbb{CU}_{global}
\end{gather}
\noindent whereby $N$ denotes the total number of abstract services.
As mentioned, there is no guarantee that satisfying the local parts at component level can lead to global satisfaction. However, it is easy to see that a violation of a global constraint is contributed by some (or all) of the component services selected, even though their local constraints may have been satisfied.

\section{Temporal Debt-Aware Utility Model}
\label{sec:debtmodel}
In the \emph{Modeling} stage of \approach, we propose temporal debt-aware utility model, a notion derived from technical debt metaphor~\cite{cunningham1992wycash}, that quantifies the long-term benefit of each service component that support a composition plan. To this end, we adopt the notion of principal and interest~\cite{tom2013exploration,ampatzoglou2015financial,alves2016identification} to analyze the debt values related to a single component service that is feasible. Built on the concept of two level constraints and their different strictness, a debt can quantify each feasible component service's local contribution to the overall debt at the global level over a period of time.

\subsection{Modeling Temporal Debt Value}

\subsubsection{Principal}

The principal, denoted as $P_{c_{xy}}$, is the one-off cost of the processes on adapting a component services $c_{xy}$. It can be calculated as:

\begin{gather}
P_{c_{xy}} = O_{c_{xy}} \times C_{com} 
\end{gather}

Suppose that the actuation process for adding a component service requires an overhead of 5 seconds (denoted as $O_{c_{xy}}$) and the execution cost of computing resource is \$ 0.005 per second (denoted as $C_{com}$), then it takes a principal as 5 $\times$ 0.005 = \$ 0.025. Note that $P_{c_{xy}}$ here is a normalized value in the range of $[0,1]$, based on the lower/upper bounds of the possible execution cost and composition time. The $O_{c_{xy}}$ can be easily known by analyzing the time for previous rounds of composition. Alternatively, it can be obtained via profiling the service broker, as what we have done in this work. 

\subsubsection{Accumulated interest} 

Over time, interests can be accumulated due to continuous constraint violations. Since the local constraints are hard, there will be no interest incurred directly at this level. However, because we model the global constraints as the soft ones, any violation of a global constraint is contributed by the component services at the local level, even if the local constraint has been satisfied. In particular, according to Equation 3 and 4, over a period of time, any possible violation of a global constraint would be contributed by all component services that have local utilization/latency worse than the global constraint, which causes potential interest. With this in mind, the accumulated interests of a component service $c_{xy}$ between timestep $n$ and $m$ can be modeled as:


\begin{gather}
I_{n,m,c_{xy}}=\alpha_{n,m,c_{xy}} + \beta_{n,m,c_{xy}}
\end{gather}
\noindent and
\begin{gather}
\alpha_{n,m,c_{xy}}= \sum_{t=n}^{m} (\varmathbb{CU}_{global} -  \varmathbb{U}_{c_{xy}}), \text{ } \forall t \text{ } \stackrel{\bullet}{\equiv} \text{ } \varmathbb{CU}_{global} \geq  \varmathbb{U}_{c_{xy}}  
\end{gather}
\begin{gather}
 \beta_{n,m,c_{xy}}=\sum_{t=n}^{m} (\varmathbb{L}_{c_{xy}} - \varmathbb{CL}_{global}), \text{ } \forall t \text{ } \stackrel{\bullet}{\equiv} \text{ } \varmathbb{L}_{c_{xy}} \geq  \varmathbb{CL}_{global}
\end{gather}

\noindent whereby $\stackrel{\bullet}{\equiv}$ represents `such that'. Hence, $\alpha_{n,m,c_{xy}}$ and $\beta_{n,m,c_{xy}}$ consider only those timesteps between $n$ and $m$, at which contribution to the possible violation of a global constraint exists. In particular, these equations guarantee that $\alpha_{n,m,c_{xy}} \geq 0$ and $\beta_{n,m,c_{xy}} \geq 0$.

It is easy to know that in general, if $\alpha_{n,m,c_{xy}} = 0$ and $\beta_{n,m,c_{xy}} = 0$, which means $c_{xy}$ does not contribute to any possible global violation at all, then the overall accumulated interest for $c_{xy}$ over a period of time is 0. Otherwise, the interest, incurred by the contribution to the possible violation of either global utilization or latency constraint (or both), would be part of the debt. For example, when $\varmathbb{CU}_{global} = 0.9$ and $\varmathbb{CL}_{global} = 0.7$, at a particular timestep $t$, a feasible component service has utilization and latency of $\varmathbb{U}_{c_{23}}=0.7$ and $\varmathbb{L}_{c_{23}}=0.85$, respectively. In this case, for any possible violation of the global utilization and latency constraint at this timestep, $c_{23}$ would contribute a total of $I_{t,t,c_{23}}=0.9-0.7+0.85-0.7=0.35$ interest (and thus part of the debt) to cause the violations. The overall interest over a range of timesteps would be the sum of the interest incurred by the above case under each timestep.

\subsubsection{Connecting debt and utility} 

Finally, we calculate the debt for a feasible component service between timestep $n$ and $m$ as:

\begin{gather}
D_{n,m,c_{xy}} = P_{c_{xy}} + I_{n,m,c_{xy}}
\end{gather}

Since both $P_{c_{xy}}$ and $I_{n,m,c_{xy}}$ are normalized or naturally sit between $[0,1]$, the numeric stability can be improved. Drawing on the above, we then be able to obtain a debt-aware utility score ($S_{n,m,c_{xy}}$) for $c_{xy}$ between $n$ and $m$, defined as:

\begin{gather}
S_{n,m,c_{xy}}=\sum_{t=n}^{m} {\varmathbb{U}_{c_{xy}}} - \sum_{t=n}^{m} {\varmathbb{L}_{c_{xy}}} - D_{n,m,c_{xy}}
\end{gather}
A larger $S_{n,m,c_{xy}}$ implies that the component service $c_{xy}$ is more likely to contribute to the satisfaction of global constraints in the long term. Here, it is clear that we will accept certain debt, as long as it can be paid back by achieving better overall utility across the timesteps considered. In this way, during the reasoning process, \approach~is able to quantify the long-term benefit of each feasible component service over a range of timesteps, based on which enabling better informed reasoning.
\subsection{Time-Series Workload Prediction}

Predicatively analyzing debt is not uncommon for managing technical debt in software development~\cite{cunningham1992wycash}. Often, the fact of whether a debt can be paid off depends on the present and future values of the debt ~\cite{buschmann2011pay,snipes2012defining}. This is also an equivalent and important concept in our research, and therefore we seek to predict the future workload of the component services, which in turn, enabling informed reasoning of long-term benefit during self-adaptation.

In \approach, we use Autoregressive Fractionally Integrated Moving Average model (ARFIMA)~\cite{xiu2007empirical}, a widely used time-series model, to predict the workload of each abstract service. It is chosen over its counterparts (e.g., ARMA) because it handles a time-series with long memory pattern well.



Accordingly, for each abstract service that is realized by a component service, we prepared the data at each time point to contain a number of observed requests, which would be used by the ARFIMA to predict the likely requests workload for a future timestep. The general expression of ARFIMA ($p$, $d$, $q$) for the process $X_t$ is written as:

\begin{gather}
\Phi(B)(1-B)^d X_t = \Theta (B)\varepsilon_t
\end{gather}

\noindent where $(1-B)^d$ is the fractional differencing operator and the fractional number $d$ is the memory parameter, such that $d \in (- 0.5, 0.5)$. The operator $B$ is the backward shift operator. For this, we have $\Phi(B)= 1-\phi_1B - \phi_2B^2 - ... -\phi_pB^p$ is the autoregressive polynomial of order $p$ and $\Phi(B)= 1+\theta_1B + \theta_2B^2 + ... +\theta_qB^q$ is the moving average polynomial of order $q$. $BX_t=X_{t-1}$ and $\varepsilon_t$ represent the white noise process.  

In Section~\ref{sec:exp-setup}, we will explain how and what tools we use to determine the values of the parameter $p$, $d$ and $q$.



\section{Debt-Aware Two Levels Constraint Reasoning}
\label{sec:reasoning}

Drawing on our formalization of soft/hard constraints at two levels, along with the proposed temporal debt-aware utility model, we design a simple yet efficient reasoning algorithm for self-adapting service composition in the \emph{Reasoning} stage. In a nutshell, once violation on local constraints is detected, the algorithm has two main functions that are run in order:

\begin{enumerate}
    \item \textsc{\underline{Identification:}} In this function, we firstly identify which are the component services that violate the local constraints, as this was what triggered the adaptation. Then, the identified infeasible component services would need to be replaced, as they also contribute to the likely violation of the global constraint(s). It is possible that all component services need to be replaced.
    
    \item \textsc{\underline{Search:}} Once we identify the set of abstract services whose component service needs a replacement, this function works on each individual abstract service. The aim is to search for the best feasible component service for each identified abstract service, such that it satisfies the local constraint\footnote{Given that the local constraint is specified at the local level, there will be at least one readily available component service to satisfy such constraint at a particular timestep, or otherwise the constraint may be too strong and needs to be relaxed.} while having the best long-term debt-aware utility, over all timesteps up to the future timestep $m$ (Equation 10). As a result, the newly selected component services would less likely to cause local/global constraint violation in the future.
\end{enumerate}

Each of the key steps are discussed in details as follows.

\subsection{Identifying Infeasible Component Services}

As mentioned, since the constraint at local level is hard, the \textsc{Identification} function is designed to filter all the service components  that are `working fine'. In fact, this steps is an effective way to reduce the search space, as only the problematic component services that violates the hard constraints are considered. These infeasible component services can actually contribute to the global constraint violation, if any. 

The corresponding algorithmic procedure has been illustrated in Algorithm 1. As can be seen, the returned result is a set, denoted as $S_{inf}$, that contains every abstract service (i.e., $a_x$) whose component service becomes infeasible at the current timestep $n$.

\begin{algorithm}[t!]
\small
\caption{\textsc{Identification}}
\SetAlgoLined
\DontPrintSemicolon
\textbf{Input}: \textit{ $S$: Set of selected component services and their abstract services at current timestep $n$} \\
\textbf{Output}: \textit{ $S_{inf} \leftarrow \varnothing$: Set of abstract services whose component service needs a replacement} \\
\For{$\forall c_{xy} \in S$}{

\If{($\varmathbb{L}_{c_{xy}} > \varmathbb{CL}_{c_{xy}}$ \textbf{or}  $\varmathbb{U}_{c_{xy}} < \varmathbb{CU}_{c_{xy}}$)}{

$S_{inf} \leftarrow a_{x}$\\

} 
}
\textbf{return} $S_{inf}$
\end{algorithm}

\subsection{Searching for the Best Long-term Debt-Aware Utility}

The special design in the \textsc{Search} function is that, instead of having to examine every combination of the service composition globally, we only search for the component service with the highest long-term debt-aware utility for each identified abstract service independently.

This is because, according to Equation 10, the problem of searching the highest long-term debt-aware utility (between timestep $n$ and $m$) for the entire service composition can be defined as follow:
\begin{gather}
\mathbf{argmax}\sum_{x=1}^{Z} S_{n,m,c_{xy}}
\end{gather}

\noindent whereby $Z$ is the total number of abstract services whose component service need a replacement. Clearly, this is a typical linear programming problem, in which achieving the best utility of the service composition is equal to finding the optimal value of each $S_{n,m,c_{xy}}$. From Equation 10, we know that the best $S_{n,m,c_{xy}}$ is solely equivalent to the highest debt-aware utility from all the feasible component services of the $x$th abstract service. In other words, the highest $S_{n,m,c_{xy}}$ can be searched on each abstract service locally, in order to have the highest utility for the service composition globally. With this consideration, our reasoning algorithm decomposes the problem and reduces the search complexity from $O(Y^X)$ (when all combinations need to be searched at the global level) down to $O(Y \times X)$, where $X$ is the number of problematic abstract service, each with $Y$ feasible component services\footnote{$Y$ may differ for different abstract services, but in this example we assume that same as our aim is merely to intuitively illustrate the reduction of complexity.}.

The corresponding algorithmic procedure has been illustrated in Algorithm 2. Specifically, suppose that the $S_{inf}$ has been found by Algorithm 1, and that the current timestep is $n$ and we are interested in the debt up to a given timestep $m$ in the future, there are three important steps:

\begin{enumerate}
    \item From line 4 to 14, for each problematic abstract service $a_x$, we firstly construct an ordered list of vectors denoted as $M_x$. Each vector in $M_x$ has a size of $m-n$ and it contains all the feasible component service for $a_x$ under every particular timestep between $n$ and $m$.
    \item From line 15 to 20, for each $M_x$, we find the largest timestep $m_x$ since $n$ such that there is at least one feasible component service that satisfies the local constraint on every timestep between $n$ and $m_x$. Next, we use the smallest $m_x$ across all $M_x$ to serve as the new $m$. This process ensures that all problematic abstract services would have at least one component service which can be treated as feasible on all timesteps considered. Here, since there is at least one feasible component service for a particular timestep, the worst case would be $m=n+1$.
    \item From line 21 to 24, for each $a_x$, we find the set of feasible component services ($S_x$) that satisfy the local constraints on every timestep between $n$ and $m$. The \textsc{SearchUtility} function searches locally on the set $S_x$, and returns the one with the highest $S_{n,m,c_{xy}}$ as part of the composition plan. Note that, \textsc{SearchUtility} can be realized by any search algorithm, e.g., exhaustive search or stochastic search like Genetic Algorithm. Since in this work the $S_x$ has been reduced to a computationally tractable size, we simply apply an exhaustive search.
\end{enumerate}

As the global constraints are soft, the reasoning algorithm has never explicitly used them to act as caps or thresholds for the search (like what we did for the hard local constraints), but the global constraints, along with their potential violations contributed by the component services, are implicitly embedded in the debt-aware utility model. In this way, we aim to mitigate the problem of being over-optimism on the global constraint, while at the same time, promoting larger chance to satisfy the global constraint in the long term.



\begin{algorithm}[t!]
\small
\caption{\textsc{Search}}
\SetAlgoLined
\DontPrintSemicolon
\textbf{Input}: \textit{$R_{x}$: The set of possible component services for the $x$th abstract service}\\ 
\textit{ $S_{inf}$: Set of abstract services whose component service needs a replacement}\\
\textbf{Output}: \textit{$S_{optimal}$: Service composition plan with the optimal long-term debt-aware utility between current timestep $n$ and the future timestep $m$}\\
\For{$\forall a_x \in S_{inf}$}{

\tcc{$M_{x}$ denotes the ordered list of vectors of the feasible component services for the $x$th abstract service at every timestep from $n$ to a future timestep $m$}

\tcc{$\mathbf{S}_{x,t}$ denotes the vector of the feasible component services for the $x$th abstract service at timestep $t$}

$M_{x} = \{\mathbf{S}_{x,n}, \mathbf{S}_{x,n+1},...,\mathbf{S}_{x,m} \}$

\For{$\forall c_{xy} \in R_{x}$}{

\For{$t \leftarrow n+1$ \textbf{to} $m$} {

\If{($\varmathbb{L}_{c_{xy}} \leq \varmathbb{CL}_{c_{xy}}$ \textbf{and}  $\varmathbb{U}_{c_{xy}} \geq \varmathbb{CU}_{c_{xy}}$)}{

$\mathbf{S}_{x,t} \leftarrow c_{xy}$
}
}

}

$M \leftarrow M_{x}$

}

\For{$\forall M_{x} \in M$}{
\tcc{According to $M_{x}$, the function \textsc{getLargestFeasibleStep} returns the largest timestep $m_x$ from $n$ such that there is at least one component service that satisfies the local constraint on every timestep between $n$ and $m_x$}
$m_x = $ \textsc{getLargestFeasibleStep($M_{x}$)}\\
\If{$m_x < m$}{
$m = m_x$
}

}

\For{$\forall M_{x} \in M$}{
\tcc{According to $M_{x}$ and the new $m$, the function \textsc{getFeasibleServices} returns the component services that satisfy the the local constraint on every timestep between $n$ and $m$}
$S_x = $ \textsc{getFeasibleServices($M_{x}$,$m$)}\\
\tcc{Function \textsc{searchUtility} returns the component service with the highest $S_{n,m,c_{xy}}$ for $a_x$}
$S_{optimal} \leftarrow$ \textsc{searchUtility}($S_{x}$,$n$,$m$)

}

\textbf{return} $S_{optimal}$
\end{algorithm}

\section{Evaluation}
\label{sec:exp}

To evaluate \approach, we design experiments to assess the performance of our technique on self-adapting service composition by means of comparing it with the state-of-the-art approaches. In particular, we aim to answer the following research questions (RQs):

\begin{itemize}

 \item \textbf{RQ1:} Can \approach~achieve better global utilization and latency than the state-of-the-art approaches? If so, which parts contribute to the improvement?
 \item \textbf{RQ2:} Is \approach~more sustainable than the state-of-the-art approaches?
    \item \textbf{RQ3:} What is the running overhead of the reasoning process in \approach~comparing with the others?
\end{itemize}

\subsection{Experimental Setup}
\label{sec:exp-setup}

Our experiments have used a commonly applied service-based system~\cite{DBLP:journals/infsof/ChenLY19,DBLP:conf/gecco/0001LY18,DBLP:conf/icpads/KumarBCLB18} with 10 abstract services, each of which has 10 possible component services to be selected. Without considering reduction, the system would have a search space of $10^{10}$ possible composition plans for self-adaptation. All the values of latency and throughput capacity for the component services are randomly chosen from the WS-DREAM dataset~\cite{zheng2012investigating}.


To emulate realistic workload for each abstract service that is realized by a component service, we extracted the FIFA98 trace~\cite{arlitt2000workload} for the length of 6 hours with 7200 timesteps, which forms the workload dataset. Such a workload trace is used on all the different workflows of service composition. We pre-processed the first four hours of workload trace as the samples for training the time-series prediction model, while the remaining two hours of workload data, which equals to 7200 seconds, is used for testing the accuracy. In \approach, we feed the training data into the ARFIMA, which is implemented using the
\texttt{arfima} package~\cite{veenstra2015package} and the \textsc{fdGPH} function in R~\cite{maechler2019package}. The values of $p$, $d$ and $q$ are also identified therein.


%

\begin{table}[t!]
\caption{Parameter settings}
\label{tb:para}
\begin{tabular}{lc}
\toprule
\textbf{Parameter}&\textbf{Value}\\ 
\midrule
$\varmathbb{CL}_{c_{xy}}$: local latency constraint per request& 0.09s\\
$\varmathbb{CL}_{global}$: global latency constraint per request& 1s\\
$\varmathbb{CU}_{c_{xy}}$: local utilization constraint& 0.8\\
$\varmathbb{CU}_{global}$: global utilization constraint& 0.9\\
$C_{com}$: cost of computing resource& \$0.0025\\
$m$: future timestep $m$ from current timestep $n$& $n+5$\\
\bottomrule
\end{tabular}
\end{table}

Table~\ref{tb:para} shows the parameter settings of the SLA used in the experiments, including the executing resource of selecting a component service ($C_{com}$), the local and global constraint for latency ($\varmathbb{CL}_{c_{xy}}$ and $\varmathbb{CL}_{global}$) and utilization ($\varmathbb{CU}_{c_{xy}}$ and $\varmathbb{CU}_{global}$). For simplicity of exposition, we have set the same local constraint for all abstract services. All the settings above have been tailored to be reasonable throughout the experiments.


All experiments were carried out on a machine with Intel
Core i7 2.60 GHz. CPU, 8GB RAM and Windows 10.

\subsection{Comparative Approaches}

To answer all the RQs, we examine the performance of \approach~against the following approaches:

\begin{itemize}
    \item[---] \textbf{Two Level Hard Constraints Approach (TLHCA):} This is similar to \approach, which differs only on the way about how the strictness of the two levels constraints is formulated. \texttt{TLHCA} assumes that both local and global constraints are hard, and thereby in the reasoning algorithm (Algorithm 2), when the final composition plan violates the global constraint (for every timestep between $n$ and the newly defined $m$) then we examine whether all abstract services have been considered in this run. If not, we then rerun the algorithm with consideration that all the abstract services are subject to replacement; if all abstract services has been considered but the global constraint(s) is still violated, we would have no choice but to trigger the adaptation. Here, the adaptation is triggered based on both local and global constraint violations. This approach follows the existing work~\cite{ardagna2005global} that makes the same formulation, and by this mean, we aim to examine the usefulness of formulating the global constraints as the soft ones.
    
    \item[---] \textbf{Debt-Oblivious Approach (DOA):} This is a similar copy of \approach~but without the temporal debt-aware utility model. Instead, \texttt{DOA} assumes the predicted utility of the aggregated latency and utilization, i.e., Equation 10 without the debt, which is then used in the reasoning algorithm to find the composition plan for self-adaptation. Such a predicted approach has been used in existing work~\cite{dai2009qos}, and \texttt{DOA} helps us to examine the effectiveness of incorporating debt information for achieving long-term benefit in self-adaptation.
    
    \item[---] \textbf{Region-Based Composition (RBC)} This is an implementation of a state-of-the-art approach, proposed by Lin et al.~\cite{lin2010design}, that relies on regions, where for each abstract services, the component service is selected according to its region. Each of these regions are clustered based on the historical utilization and latency of the component services. Here, the adaptation is triggered based on global constraint violations only. \texttt{RBC} is chosen as it is one of the most widely known representative approaches for dynamic service composition.
\end{itemize}

%
%

\subsection{Metrics}

We leverage the following metrics to assess the results:
\begin{itemize}
    \item[---] \textbf{Global utilization:} This is the value calculated by Equation~\ref{eq:globalutil} for each timestep.  
    \item[---] \textbf{Global latency:} This is the value calculated by Equation~\ref{eq:globallatency} for each timestep.  
    \item[---] \textbf{Accumulated debt:} Since the interests are accumulated, so does the debt. A lower debt means that component services, which are less likely to contribute to global constraint violation in the long term, are preferred. Therefore, we measure the accumulated debt of the service composition from the beginning to the timestep $t$ using:
    \begin{gather}
D_{1,t}=\sum_{x}\sum_{y} D_{1,t,c_{xy}}
\end{gather}
\item[---] \textbf{Sustainability score:} We measure sustainability as follows:
  \begin{gather}
Score_{n,m}={1\over V} \times ({{S_{n,m} - S_{min,n,m}} \over { {S_{max,n,m} - S_{min,n,m}}}}+1)
\end{gather}
\noindent whereby $S_{n,m}={\sum_{x=1}^{Z} S_{n,m,c_{xy}}}$; $n=1$ and $m=7200$; $Z$ is the total number of abstract services; $V$ is the total number of local and global constraint violations. $S_{min,n,m}$ and $S_{min,n,m}$ are the lower and upper value among all approaches. $Score_{n,m} \in [1,2]$ and a higher value means that the adaptations would generate more benefits in general when mitigating each constraint violation.
    \item[---] \textbf{Running time:} This is the required running time for the reasoning process to produce a composition plan.  
\end{itemize}

Whenever overall results are reported, we use the pairwise version of the Kruskal Wallis test ($\alpha=.05$)~\cite{kruskal1952use} and $\eta^2$ value~\cite{cohen2013statistical} to measure statistical significance and effect size, respectively.

\subsection{RQ1: Performance of \approach}

\begin{figure}[t!]
    \centering
\includegraphics[width=0.7\columnwidth]{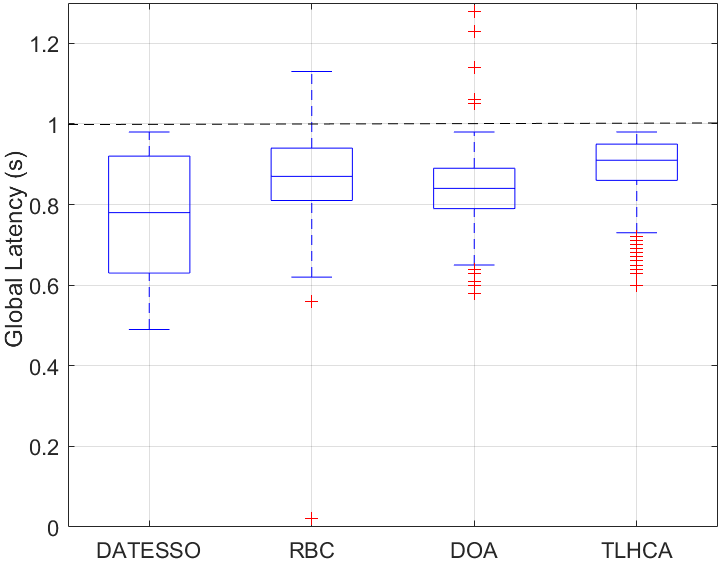}
    \caption{Global latency yield by all approaches over 7200 timesteps (Comparisons between \approach~and others are statistically significant ($p<.05$) and with large effect size)}
    \label{fig:latency}
\end{figure}

Figure~\ref{fig:util} and~\ref{fig:latency} respectively illustrate the global utilization and latency for all approaches and timesteps. As can be seen, the comparison between \approach~and any other three are statistically significant with large effect size. In particular, when comparing with \texttt{RBC}, \approach~achieves much better utilization and latency overall, while at the same time, it has smaller variance than \texttt{RBC}.

To better understand which of our contributions in \approach~enable such improvement, we firstly compare it with \texttt{TLHCA} and \texttt{DOA}. As shown in the boxplots, we see that \approach~achieves much better utilization and smaller variance. For latency, \approach~is slightly more deviated, but provides overall better result. This has proved that, in general, the formalization of two levels constraints with different strictness can help to improve self-adaptation performance. Next, we compare \approach~with \texttt{DOA}, for which we see that again, \approach~achieves generally better and more stable results on utilization and latency. This evidences that the predicted debt model can provide more benefit than simply having a predicted model based solely on utilization and latency. 

Remarkably, \approach~achieves full satisfaction for the global constraint on latency and satisfy that of utilization for majority of the cases, which are generally superior to the other three. Therefore, for \textbf{RQ1}, we conclude that:



\begin{figure}[t!]
    \centering
\includegraphics[width=0.7\columnwidth]{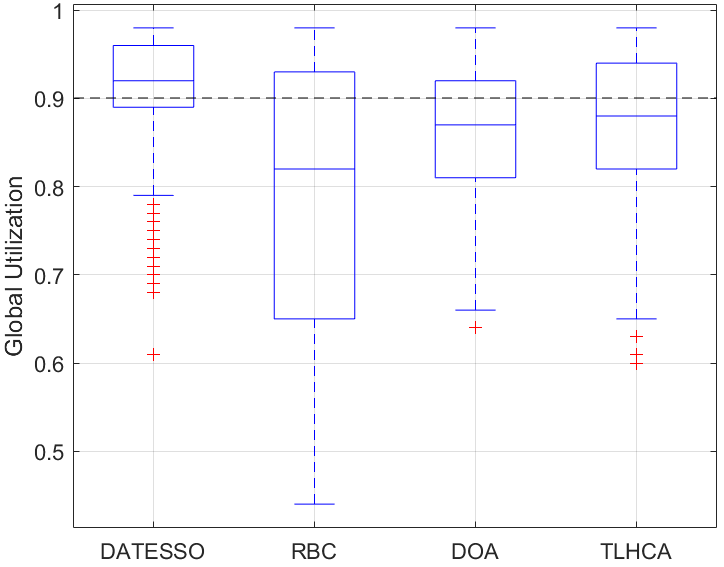}
    \caption{Global utilization yield by all approaches over 7200 timesteps (Comparisons between \approach~and others are statistically significant ($p<.05$) and with large effect size)}
    \label{fig:util}
\end{figure}

\begin{tcolorbox}[breakable,left=5pt,right=5pt,top=5pt,bottom=5pt] 
\underline{\textbf{Answering RQ1:}} \approach~is more effective than the state-of-the-arts in terms of the utilization and latency, with better satisfactions. Both the design of formalizing global constraints as the soft ones and the temporal debt-aware utility model have contributed to the improvement.
\end{tcolorbox}

\subsection{RQ2: Sustainability of \approach}

\begin{figure}[t!]
    \centering
\includegraphics[width=0.7\columnwidth]{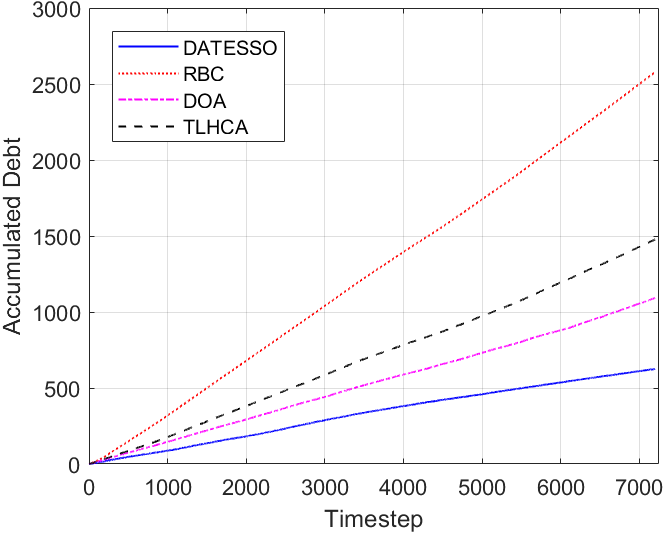}
    \caption{Total debt accumulated by all approaches over 7200 timesteps}
     \label{fig:debt}
\end{figure}

We now assess the sustainability of adaptation achieved by using the accumulated debt and sustainability score. Figure~\ref{fig:debt} shows the accumulated debt, in which we see that all approaches have accumulated debt in a linear and steady manner. However, clearly, \approach~results in significantly less debt than the other three as it accumulates overtime, suggesting that \approach~favours component services that is less likely to contribute to global constraint violation in the long term.  

Table~\ref{tb:subs} shows the sustainability scores for all approaches. As can been seen, despite that \approach and \texttt{DOA} have similar total number of constant violations, \approach~has achieved the best $Score_{n,m}$ value among others. This implies that the adaptations in ~\approach~would create the greatest benefit in mitigating per violation. All the above conclude that:
\begin{tcolorbox}[breakable,left=5pt,right=5pt,top=5pt,bottom=5pt] 
\underline{\textbf{Answering RQ2:}} \approach~is more sustainable than the other three, as it has less accumulated debt and with the highest sustainability score. This means that \approach~favors more reliable component services in the long term, and that it offers greater benefit when dealing with each violation overall. 
\end{tcolorbox}

\begin{table}[t!]
\caption{Sustainability scores}
\label{tb:subs}

\begin{tabular}{cccc}\toprule
\textbf{Approach}&$\sum_{x=1}^{Z}S_{n,m,c_{xy}}$&$V$&$Score_{n,m}$\\ \midrule
\approach&417.10&113&.0177\\
\texttt{RBC}&$-$3146.66&187&.0053\\
\texttt{DOA}&$-$910.61&102&.0160\\
\texttt{TLHCA}&$-$1478.67&133&.0110\\
\bottomrule
\end{tabular}
\end{table}

\subsection{RQ3: Running Time of \approach}

\begin{figure}[t!]
    \centering
\includegraphics[width=0.7\columnwidth]{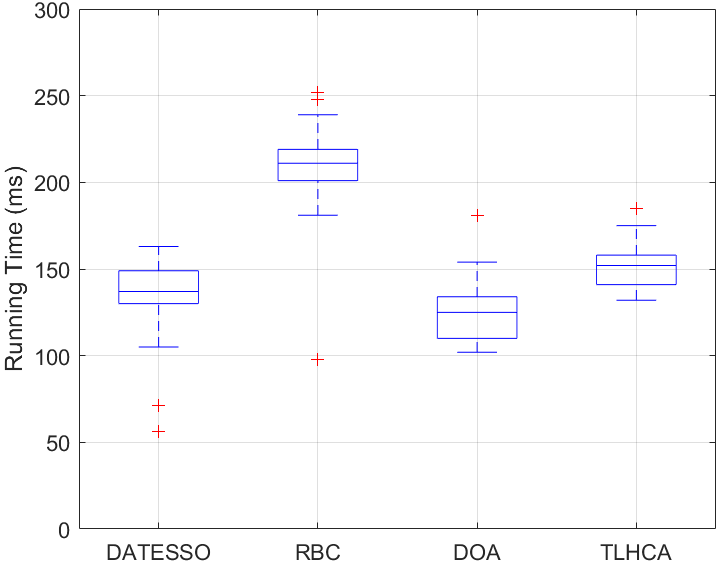}
    \caption{Running time on all approaches (Comparisons between \approach~and others are statistically significant ($p<.05$) and with large effect size, except for \texttt{DOA})}
       \label{fig:rt}
\end{figure}

Figure~\ref{fig:rt} illustrates the running time for all approaches. We can clearly see that \texttt{RBC} is the slowest due to the region based algorithm. \texttt{TLHCA} is the 2nd slowest because of the frequent need of replacing all component services. Since \approach~and \texttt{DOA} differ only on whether having the debt calculation, they have similar running overhead ($p>.05$) but are significantly faster than the others. This is because only the problematic abstract services, along with those component services that satisfy all considered timesteps, are involved in the actual search, which reduces the search space. However, as we have shown, \approach~offers much better performance and sustainability than \texttt{DOA}. In summary, we have:
\begin{tcolorbox}[breakable,left=5pt,right=5pt,top=5pt,bottom=5pt] 
\underline{\textbf{Answering RQ3:}} \approach~and \texttt{DOA} both have similar running time, but they are faster than the other two. 
\end{tcolorbox}

\section{Threats to Validity}
\label{sec:tov}

\textit{Threats to construct validity} can be related to the metric and evaluation methods used. To mitigate such, we use a broad range of metrics for evaluating different aspects of \approach, including utilization, latency and sustainability etc. To examine the effectiveness of each contribution, we have compared \approach~with specifically designed approaches, i.e., \texttt{TLHCA} and \texttt{DOA}, in addition to a direct implementation of existing work (\texttt{RBC}). Further, we plot all the data points in a trace, and applied statistical test and effect size interpretation when it is difficult to show all the data points.


\textit{Threats to internal validity} can be mainly related to the value of the parameters for \approach. Particularly, the setup has been designed in a way that it produces good trade-off between the quality and overhead. They have been shown to be reasonable following preliminary runs in our experiments. The future timestep $m$ is also specifically tailored and the used value tends to be sufficient. However, it is worth noting that the actual future timesteps to use is updated dynamically depending on whether there is a feasible component service that satisfies all considered timesteps.



\textit{Threats to external validity} can be associated with the environment and the dataset that are used in the experiment. To improve generalization, we apply commonly used service-based system~\cite{DBLP:journals/infsof/ChenLY19,DBLP:conf/gecco/0001LY18,DBLP:conf/icpads/KumarBCLB18}, whose data is randomly sampled from the real-world WS-Dream dataset~\cite{zheng2012investigating}, along with the FIFA98 workload trace~\cite{arlitt2000workload}. A more comprehensive evaluation on different dataset and more complicated structures are parts of the future work.

\section{Related Work}
\label{sec:rw}


Self-adapting service composition is certainly not new for research on service-based systems. Among others, Lin et al.~\cite{lin2010design} and Li et al.~\cite{li2010towards} rely on region-based composition, in which an expand region algorithm is proposed to identify the region of each component service, which forms a reduced search space. Dai et.al.~\cite{dai2009qos} leverage time-series prediction on workload when reasoning about the self-adaptation. \citeauthor{DBLP:journals/infsof/ChenLY19}~\cite{DBLP:conf/gecco/0001LY18,DBLP:journals/infsof/ChenLY19} seed the multi-objective evolutionary algorithms to accelerate the reasoning process of service composition. The commonality of the above work is the fact that they all assume both local and global constraints are hard ones during reasoning, which can be over-optimistic. Therefore, they can easily lead to the situation of `no satisfactory composition plans found'. \approach, in contrast, formalizes the global one as soft constraint, which mitigates the issues of over-optimism and also reward some plans that may temporarily cause global violation, but tends to be more sustainable with larger long-term benefit.

Technical debt has been studied in service composition~\cite{alzaghoul2013cloudmtd,skourletopoulos2016quantifying} and in a wider context of self-adaptive systems~\cite{DBLP:conf/wosp/0001BWY18}. For example, \citeauthor{DBLP:conf/wosp/0001BWY18}~\cite{DBLP:conf/wosp/0001BWY18} have used technical debt as a metaphor to model the problem of \emph{to adapt or not to adapt}. To resolve such, an online classifier, combined with debt calculation, is proposed. However, the above work does not explicitly consider time-varying and accumulated properties of the debt.


In summary, the key additions in \approach~are that

\begin{itemize}
    \item \approach~formalizes different strictness for the two levels constraint in service composition.
    \item \approach~makes use of a new debt model that was designed based on the different strictness of the two levels constraints and time-series prediction. It is therefore temporal, capable of quantifying accumulated debt and tailored to the problem context.
    \item Drawing on the above, \approach~proposes to leverage a simple but effective and efficient reasoning algorithm that reduces the search space and focuses on the long-term benefits of self-adaptation.
\end{itemize}

The benefits of all the above contributions have been experimentally demonstrated in Section~\ref{sec:exp}.

\section{Conclusion}
\label{sec:con}

In this paper, we propose a debt-aware two level constraint reasoning approach, dubbed \approach, for self-adapting service composition. \approach~formalizes the global constraints as the soft ones while leaving only the local ones as hard constraints. Such formalization is then used to built a temporal debt-aware utility model, supported by time-series prediction. The utility model, together with the different strictness of the two level constraints, enable us to design a simple yet efficient and effective reasoning algorithm in \approach. Experimental results demonstrate that \approach~is more effective that state-of-the-art in terms of utilization, latency and running time, while being about to make each self-adaptation more sustainable.

In future work, we seek to extend \approach~for better synergy between Software Engineering and Artificial Intelligence driven self-adaptation~\cite{DBLP:journals/corr/ChenFBLYME14,DBLP:journals/computer/LewisCFGCBTY15,DBLP:journals/pieee/Chen20}, particularly on stochastic multi-objective search algorithms which have been shown to provide promising results on scenarios with complex trade-off surface for self-adaptive software systems~\cite{DBLP:journals/tosem/ChenLBY18,DBLP:journals/tsc/ChenB17,DBLP:journals/computer/ChenB15,DBLP:journals/csur/ChenBY18,DBLP:journals/corr/abs-2001-08236,DBLP:conf/icse/Li0Y18,DBLP:journals/jss/SobhyMBCK20,DBLP:journals/corr/abs-2002-09040,icse2020-empirical}. Online learning based prediction on the satisfaction of local/global constraints~\cite{DBLP:conf/icse/Chen19b,DBLP:conf/icse/ChenB13,DBLP:conf/ucc/ChenBY14,DBLP:journals/tse/ChenB17} is also part of our ongoing research agenda.

\bibliographystyle{ACM-Reference-Format}
\bibliography{refdatabase}


\end{document}